\documentstyle[epsf,12pt]{mrsproc}
\begin{document}
\begin{center}
{\bf AB INITIO 
 CORE-LEVEL SHIFTS IN METALLIC ALLOYS  }

\vspace{0.2cm}
{Vincenzo Fiorentini,$^1$ 
	Michael Methfessel,$^2$ and Sabrina Oppo$^1$}\\  
\vspace{0.3cm}
{\it (1)\, INFM -- 
Dipartimento di Scienze Fisiche, Universit\`a di
Cagliari, Italy \\ 
(2)\, Institut f\"ur Halbleiterphysik, P.O. Box 409, 	 D-15204
Frankfurt/Oder, Germany }

\end{center}

\vspace{-1cm}
\begin{abstract}
\vspace{-0.4cm}
{\small  Core-level shifts and core-hole screening effects in alloy formation
are studied {\it ab initio} by constrained-density-functional total-energy 
calculations. For our case study, the ordered  intermetallic alloy MgAu, 
final-state effects are essential to account for the experimental Mg 1$s$ 
shift, while they are negligible for Au 4$f$.  We explain the differences 
in the screening  by analyzing the
calculated charge density response to the core hole perturbation.}
\end{abstract}

\vspace{-0.6cm}
\section{INTRODUCTION: AB-INITIO CALCULATIONS OF  CORE-LEVEL SHIFTS}

\vspace{-0.3cm}
Core levels are in a way the fingerprints of atoms. 
Since core levels are affected by the 
environment surrounding the atom, the observation of their relative
position, {\it i.e.} of core-level {\it shifts} \cite{egel}, 
allows to analyze the charge transfer and
structural changes undergone by the atoms 
when a crystal is formed \cite{wil}, when an atom is at the
surface \cite{zangwill} 
 rather than in the bulk of a
solid, or when an alloy is formed out of two elemental
solids \cite{wert,darrah}.

In the case of alloy formation (the subject of this paper),
a major objective has been to determine
the charge transfer between the constituents. Clearly, charge
 moving from one atom to another causes potential shifts at the 
atomic sites, which modify the core level binding energies.
However, the relation between these shifts and some measure
of the charge transfer is not unambiguous. In the simplest form
of the ``potential model'', the change in the average
potential felt by a core electron in a binary compound 
with a valence charge transfer $\Delta q$ is taken as
\begin{equation}
\vspace{-0.1cm}
   \Delta V = \Delta q\, (k-M) \label{pm}
\end{equation}
where $k$ and $M$ describe the intra- and inter-atom 
responses, respectively \cite{darrah}. Thus, $k$ is similar to a
Hubbard $U$ parameter and $M$ is a Madelung contribution.
These parameters can be estimated from other known quantities,
but this involves  guesswork and is made difficult by the
cancellation between $k$ and $M$. More importantly, the 
potential model of Eq.~(\ref{pm}) is only valid for 
the ``initial state'' picture, {\it i.e.} to describe the 
the positions of the core levels in the alloy and the pure metal
{\it before} a core electron is removed. To compare to
the measured binding energies, a final-state screening contribution
must be taken into account. This can be included in the formalism
only by adding  yet another parameter to the model.

It seems thus desirable to make a detailed
analysis of a typical system using a method which can quantify
the various contributions unambiguously. In the present work,
we use {\it ab-initio} density-functional total-energy
calculations to study the MgAu alloy, for which accurate
core level shifts have been measured. By studying an
 effectively isolated excited atom,
the screening of the core hole by the valence electrons is
described correctly. The core-level shift is cleanly separated  
into initial-state and final-state
relaxation contributions, which can then be checked against the
appropriate models. In addition, direct inspection of 
densities of states (DOS), on-site 
charges, and screening charge distributions
gives an understanding of the effects of alloying, and the
different screening responses to a core hole.

Ultimately, we find that the final-state contribution is very important,
as  it changes both the sign and the magnitude of the 
core-level shift on the Mg atom. The screening
of the Mg 1$s$ core hole is substantially less effective in the
alloy than in the pure Mg metal. In turn, the change in screening 
can be attributed to the
reduced Mg valence charge in the alloy. This  secondary 
consequence of the Mg--to--Au charge transfer is at least as 
important as the direct effect on the initial-state core levels. 
The initial-state  core-level shifts turn out to be compatible 
with an simple rigid-band model.

{\it Calculation of core-level shifts --}
In the measurement of core-level binding energies, an 
electron is emitted from the core state into the vacuum.  In a metal,
another electron moves in to screen the positive charge of the core
hole. The reservoir providing the screening electron is that of the
 valence electrons of the surrounding infinite crystal. Effectively
then, the 
core electron has been lifted to the Fermi level E$_{\rm F}$.
Therefore the  initial-state estimate of the core binding 
energy is the position of the core eigenvalue relative to E$_{\rm F}$ 
(not relative to the vacuum potential).  The energy relevant to 
spectroscopic observation (the measured binding energy of the core 
state including final-state \cite{mahan}
 screening effects) can be expressed as the difference in
total energy before and after emission of the core electron. 
The electronic system is neutral and in equilibrium subject to 
each prescribed core occupation, with the screening electron 
added to the valence band in the case of a core hole.

The {\it ab initio} theoretical analysis of the final-state core 
binding energies is difficult: the presence of a core hole breaks the
translational symmetry of the lattice, requiring the use of
appropriate supercells where the core hole is decoupled from its
periodic images, or of equivalent treatments. Also, it is imperative
to allow for self-consistency of the charge rearrangement around the 
core hole. The  alloy core-level shift, including 
final-state relaxation  effects,  is calculated as
\begin{equation}
    \Delta =  [E_{\rm TOT}^{\rm alloy} (q=1) - 
        E_{\rm TOT}^{\rm alloy} (q=0)] - 
[E_{\rm TOT}^{\rm const} (q=1) - 
        E_{\rm TOT}^{\rm const} (q=0)],
\label{rig} 
\end{equation}
where the total energies of supercells of the alloy and the pure
constituent elements are evaluated for one electron and no
electron ($q=1$ and $q=0$, respectively) 
promoted out of the relevant core level to the 
valence band. For an accurate DFT 
calculation, the resulting shift should be 
close to the experimental value. The initial-state estimate is obtained 
by comparing the core eigenvalues relative 
to E$_{\rm F}$ for the two systems. 
The difference of the full and initial-state results
is, by definition, the final-state screening contribution.
In this context,
 it is advantageous to use an all-electron method, which gives direct
information on core electron levels.
Note that surface properties (specifically, the work function)
 do not enter either description. This
must be the case for an acceptable model: {\em the exact core binding
energy}, expressed as the difference of two total energies, {\it  is a bulk
property}.

Our self-consistent electronic-structure and total-energy calculations
were done 
with the all-electron full-potential LMTO method \cite{fp}, within the
local approximation (LDA) to density-functional theory (DFT) \cite{gj}.
Minimization of the energy under a constrained core occupation
is rigorously justified \cite{dede} in the DFT framework: 
a self-consistent calculation under the chosen constraint
provides a variational total energy in the parameter subspace
identified by the constraint, and the rigorous definition of core
level shifts as differences of total energies can be applied. 
To calculate the final-state core-level
shifts, we perform total-energy calculations at various occupations
of the relevant core state in appropriate supercells
(see below).

{\it Connection between initial-state and exact core-level 
shifts --} 
Within DFT, one can adopt an approach similar to Slater's
transition-state concept \cite{slater}, which adds to our understanding
of final-state effects. This is based on Janak's formula \cite{janak},
which for the present purposes can be restated as
\begin{equation}
 \frac{\partial E_T}{\partial q} = E_{\rm F} - E_c (q) \equiv
\epsilon_c (q), \label{jak}
\end{equation}
where $E_{\rm F}$ is the Fermi energy, $E_c$ is the core-level
eigenvalue, and $q$ is the promoted charge.  If we assume that the
core level energy varies linearly with the occupation of the level (a
good approximation), the total-energy difference of Eq.\ref{rig} can
be expressed via trapezoidal-rule integration of Eq.\ref{jak} between
zero and unit promoted charge, as
\begin{eqnarray}
\Delta & \approx & {\textstyle \frac{1}{2}}
  \Big\{[\epsilon_c^{\rm alloy}(1) + \epsilon_c^{\rm alloy}(0)] -
[\epsilon_c^{\rm const}(1) + \epsilon_c^{\rm const}(0)]\Big\}
\label{corfs}
\\ \nonumber
 & \approx & \Delta_{\rm I} + {\textstyle \frac{1}{2} }
\Big\{[\epsilon_c^{\rm alloy}(1) - \epsilon_c^{\rm alloy}(0)] -
[\epsilon_c^{\rm const}(1) - \epsilon_c^{\rm const}(0)]\Big\}
\vspace{-0.3cm}
\end{eqnarray}
where the second line rephrases the alloy core-level shift as a
correction to the initial-state estimate $\Delta_{\rm I} =
\epsilon_c^{\rm alloy}(0) - \epsilon_c^{\rm const} (0)$.  This 
identifies the correction due to final-state screening effects as the
difference of core eigenvalue {\it drop} upon depopulation of the
level in the two environments.  The above description helps to
make contact to differences in the screening response of the two 
environments in question. In general terms, the eigenvalue of 
the depopulated core state drops by a larger amount when the valence 
electrons screen the core hole less efficiently.

Formally, the occupation of a state can be varied continuously in DFT,
which allows self-consistent calculations at arbitrary
occupation.  Therewith one can verify that the total energy
derivative (Eq.\ref{jak}) is very nearly linear in the occupation
number of the state. The second derivative of the total
energy vs. promoted charge is positive, which of course indicates that
at non-zero promoted charge 
the system is unstable towards reoccupation of the
state \cite{proto}. This, however, does not affect the variational
character of the total energy under each occupation
constraint \cite{dede}.

\vspace{-0.6cm}
\section{A CASE STUDY: M\lowercase{g}A\lowercase{u}}
\label{mgau-sec}

\vspace{-0.3cm}
We applied the technique outlined above to the core shifts of the Mg
1$s$ and Au 4$f$ levels upon formation of the binary,
CsCl-structure MgAu alloy (assumed to be ordered) 
out of  bulk Au and Mg.  
Accurate experimental data \cite{wert,darrah} exist for the core level 
shifts upon formation of this alloy.
The calculated structural parameters for MgAu, Au and Mg are given 
in Table \ref{t1}. The results for these bulk systems are of standard
DFT-LDA quality. For Mg, the fcc structure was adopted. 

{
\begin{table}[h]
\centering
\begin{tabular}{|l|ccc|}
\hline
 & a$_0$ (bohr) & B$_0$ (Mbar) &
E$_{\rm coh}$ (eV/atom)  \\
\hline
MgAu th.  & 6.09 & 1.05  & 3.70  \\
MgAu exp.   & 6.15  & ---  &  --- \\
\hline
Au th.  & 7.68 & 1.85 & 4.33 \\
Au exp. & 7.70 & 1.73 &  3.81 \\
\hline
Mg th.  & 8.38 & 0.40  & 1.69  \\
Mg exp.   & 8.46   & ---  &  --- \\
\hline
\end{tabular}
\caption[T]{\small  Structural properties of 
MgAu (CsCl structure), Au (fcc) and Mg  (fcc).
Cohesive energy do  not include spin-polarization.
The Mg experimental  lattice constant corresponds  to the experimental
volume per atom in the hcp structure.}
\label{t1}
\vspace{-0.3cm}
\end{table}}

To study the  core-hole--excited solids, we used 
 16-atom supercells for both CsCl-structure
MgAu, and fcc Mg and Au. The distance of the core hole from its
periodic images exceeds 12 bohr in all cases, and  tests show that our
values for the core level shifts are converged with respect to
cell dimension. The localization of the calculated density response
provides an {\it a posteriori} justification for the supercell 
approach. The Brillouin-zone integration was done using more than 50
irreducible special points.
Muffin-tin radii for Mg (Au) are 2.94 (2.60) bohr
in the pure metal and 2.50 (2.70) in the compound at the 
experimental lattice constants, and are scaled with the
lattice constant. All the calculations are scalar-relativistic.

As already mentioned, the core eigenvalues extracted from bulk 
calculations provide the
initial-state estimate of the core shifts (see text after Eq. 4).
An electron is then removed from each core state of
interest in each of the bulk supercells, 
and the full ``final-state'' core level
shifts are obtained as total energy differences. Using fractional
occupation, we verified the linear behaviour of $\epsilon_c$,
and the validity of the Slater transition state rule.
 
The results of the calculations in the different approximations 
are summarized in Table \ref{t2}, and compared with the experimental
values of Ref. \cite{darrah}. The full results are in very
good agreement with experiment for both cases.  The initial-state 
estimate is accurate for the Au 4$f$ shift, but it is 
incorrect in sign and magnitude for Mg 1$s$. The screening contributions 
to the shifts are thus completely different in the two cases: negligible 
for Au, very substantial for Mg. 
{
\begin{table}[h]
\centering
\begin{tabular}{|l|ccrc|}
\hline
shift &
exp &
full &
initial &
screening \\
\hline
Au 4$f$   & 0.74  & 0.73 & 0.74 & 0.01 \\
Mg 1$s$ & 0.34 & 0.25 & --0.44 & 0.69 \\
\hline
\end{tabular}
\caption[T]{\small   Mg 1$s$ and Au 4$f$ core-level 
shifts in MgAu with respect to Au and Mg bulk. All values are 
in eV. Column ``full'' is the full calculation,
column ``initial'' is the initial-state shift
 (difference of core eigenvalues
 at zero promoted charge),
column ``screening'' is the difference of the previous two, and is the
contribution due to screening effects.
Experimental data from Ref.\cite{darrah}.}
\label{t2}
\vspace{-0.3cm}
\end{table}}
This is a valuable piece of information
provided by the {\it ab initio}  calculation. 
While appropriate for Au, 
the initial-state 
approximation gives a qualitatively incorrect 
picture of the Mg 1$s$ shift.
This is due to the differences in the screening 
response at the Mg site in the alloy as compared to the elemental 
bulk. The positive shift (obtained from both full final-state calculation 
and  experiment) means that the core eigenvalue drops
more strongly when depopulated in the alloy than in elemental Mg.
So the  screening response to the core-hole perturbation
at the Mg site is less effective in the alloy than in
 Mg bulk: clearly, this could not be inferred from an initial-state or
 model estimate. 
{
\begin{table}[h]
\centering
\begin{tabular}{|l|rr|}
\hline
case &
{exp}&
{calc}\\
\hline
Au 4$f$ in Au bulk        & 85.88  & 85.78  \\
Au 4$f$ in MgAu     & 86.62  &  86.51 \\
\hline
Mg 1$s$ in Mg bulk        & 1303.20 & 1306.78\\
Mg 1$s$ in MgAu         & 1303.54  &  1307.03  \\
\hline
\end{tabular}
\caption[T]{\small   Calculated vs. measured core-level 
binding energies, in eV,
referred to the Fermi level.
Experiment: Ref.\protect\cite{darrah}}
\label{t3}
\vspace{-0.3cm}
\end{table}}
Note that even the {\it absolute} core 
binding energies (referred to $\rm E_F$, and 
obtained as total-energy differences) are in remarkable agreement 
with experiment. Errors are well below 1\%, as can be seen in Table 
\ref{t3}.  This may be not too surprising, since the core levels are 
obtained as differences of two exact (within DFT-LDA) total energies.

Also, we mention that the solid with a core hole 
is the initial-state configuration for the primary-hole recombination 
with ensuing emission of the Auger electron \cite{wei}.
The initial-state estimate of  the Mg$\rightarrow$MgAu shift
of the {\it KLL} Auger transition  extracted from 
our calculated eigenvalues in the core-hole--excited solid  
 is 1.3 eV, which compares reasonably with the experimental value of
1.0 eV \cite{darrah}. (Of course, this agreement may 
result from the cancellation of multiple-core-hole \cite{egel,wei} 
relaxations.)

{\it Initial-state shifts --}
Initial-state-based models, used to infer charge transfer from
core-level shifts, rely implicitly  on a rigid-band picture: 
a shift of the DOS of the constituents is
related to charge transfer from one species to another.
Assuming that core levels move rigidly along with the valence DOS,
an estimate of the charge transfer from species 2 to 1 is
$$\Delta q^{(1)} \equiv D_{\rm F}^{(1)}\, \Delta \epsilon_c^{(1)} =
  D_{\rm F}^{(2)}\, \Delta \epsilon_c^{(2)} 
\equiv - \Delta q^{(2)}
$$ 
where $D_{\rm F}$ is the DOS
at the Fermi level, and $\Delta \epsilon_c$ is the core level shift.
The core shift measures, via the Fermi-level DOS,
 the amount of depletion or filling of the valence DOS,
{\it i.e.} the charge transfer. Our  initial-state results
agree with this simple picture. Using calculated initial-state shifts 
and Fermi-level DOS,
we get opposite sign and about equal moduli of the
$\Delta q$'s: --0.195 electrons for Mg, and 0.199 for Au, which 
are  reasonable estimates of the charge transfer.
Of course, this picture is insufficient to account for the full
results, which contain a large screening contribution.

{\it Final-state screening contribution --}
To understand the environment  dependence of the screening,
 we examine  differences of total charge densities of the
perturbed and the unperturbed supercells of the three materials
around an excited Mg or Au atoms:
these differences are  the response or screening density. 
\begin{figure}[t]
\epsfclipon
\epsfysize=6cm 
\centerline{\epsffile{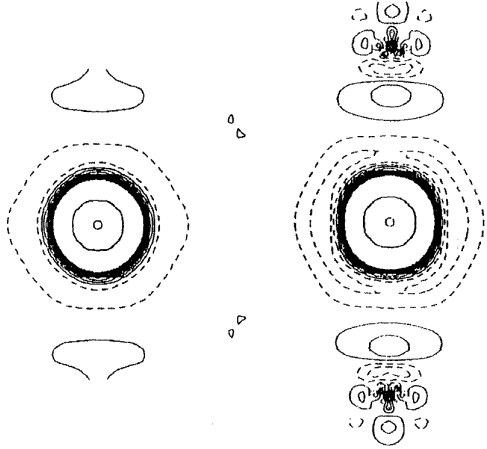}}
\caption[T]{\small
 Screening charge density for a 4$f$ core hole at the Au site 
in MgAu (left) and in Au bulk (right). All pictures drawn on the same scale.
Solid (dashed) line: positive (negative) values.}
\label{f2}
\vspace{-0.3cm}
\end{figure}
The response density of Au bulk to the Au
4$f$ core excitation is very localized, and due  mostly
to the transition-metal--like screening of the 5$d$ electrons 
(right panel in Fig. \ref{f2}). The
Au $d$-shell contains 10 electrons both before and after the core hole
is created. Our picture for the Au screening is that the $d$ shell
moves closer to the nucleus, leaving behind a charge-depleted ring;
the screening electron from  valence  $sp$ states fills up this ring. 
In the MgAu alloy, the screening is almost identical to that in Au 
bulk, very  short-ranged and dominated by the $d$ shell
(left panel of Fig. \ref{f2}); 
accordingly, the core shift has practically no 
screening-related contribution, and the 
initial-state estimate is accurate. This is in full agreement 
with both calculation and experiment.

\begin{figure}[t]
\epsfclipon
\epsfysize=6cm 
\centerline{\epsffile{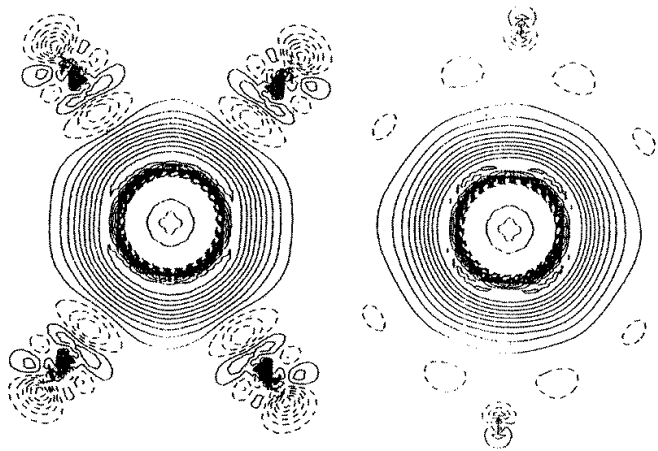}}
\caption[T]{\small
 Screening density for a 1$s$ hole at the Mg site 
in MgAu (left) and in Mg bulk (right).}
\label{f1}
\vspace{-0.3cm}
\end{figure}

As was to be expected, in bulk Mg the  response density 
to a Mg 1$s$ hole  is a roughly spherical
screening lump of $sp$ nature (Fig. \ref{f1}, right panel);
its extension is normal on the scale of the Mg-Mg interatomic distance
in the bulk,
but appreciably larger than that of the response density around Au
in its own bulk. 
Since the interatomic distances are $d_{\rm Au-Au} =
5.43$ bohr in Au, $d_{\rm Mg-Mg} = 5.92$ bohr in Mg, and $d_{\rm
Au-Mg} = 5.27$ bohr in MgAu, we expect sensitivity of
the spatially-extended screening response of Mg  to the surrounding
environment in MgAu. Indeed this is the case, as seen in 
Fig.\ref{f1}. Now, there are two basically different ways of interpreting
 the final-state effects for Mg.
{\bf First}: alloying changes the properties of the unexcited Mg atom 
(essentially, by charge transfer); only this modified screening  
ability of the Mg atom matters.
For Mg, there is less screening charge spread out
over the atomic volume than in the bulk constituent,
which  reduces the screening efficiency. The
 site-projected density of states at the Fermi level 
has also decreased (by a factor of 2), and this
again reduces the available ``mobile'' charge.
{\bf Second}: concurrently to the first mechanism,
the Mg screening charge extends far enough as to be   sensitive to  the
its neighborhood.
Charge is subtracted by the neighbors to the screening lump, making the
response less efficient.

The second explanation 
is tempting in view of the screening
response shape in Fig.\ref{f1} (left).
While in Mg bulk (right panel) only Mg atoms relatively far from
the excitation site are present, in MgAu eight polarizable Au
neigbors surround the excited Mg site at an appreciably ($\sim$10 \%)
shorter distance. Contour counting in Fig.\ref{f1}\,
already  suggests that the  screening charge at  the core-hole 
is partially depleted. This also results from  the 
integrated screening charges ${\rm
Q_{\rm scr}}$ within  spheres surrounding the core-hole
site. Assuming sphere radii of 2.48 bohr for Au and 2.68 a.u for Mg, 
one gets  ${\rm
Q_{\rm scr}}=0.840$ and 0.680 electrons for Mg in Mg bulk and MgAu,
respectively. The depletion in the alloy results in a large positive 
contribution  to the  shift, as calculated and observed.
For Au, ${\rm Q_{\rm scr}}=0.845$ and 0.821 electrons in Au bulk and 
MgAu respectively.
(Of course,   integrated charges within spheres are to some extent
 arbitrary,  and should only  be taken as qualitative indicators.)

On the other hand, while the density response is 
undoubtedly modified by the neighbors, most of
the difference in screening response comes from within 
the Mg atom.
Most of the spherically-averaged
Mg-to-MgAu difference of screening potential around the Mg site
(whose  integral gives the value of the potential shift at the 
nucleus)  is bounded within 
half interatomic distance, although it remains 
non zero outside as  well. We suggest 
that it is a combination of the two mechanisms just outlined
that causes the screening deterioration at the Mg site in MgAu, 
the bulk of it being due to pure charge transfer, with 
comparatively minor effects caused by shape and extension of  the 
screening response density.

{\it Antiscreening around Mg in MgAu --}
While possibly not  central   to the understanding of the 
screening contribution to the Mg 1$s$ shift, 
the ``antiscreening'' feature on Au neighbors visible in the Mg 1$s$
density response in Fig. \ref{f1}, right panel, is quite interesting.
We suggest that it may be interpreted as follows.  The  Friedel 
oscillation wavelength
for an electron gas having the average density of Mg 
($\overline{\rho}=0.024$  bohr$^{-3}$),  is $\lambda=3.7$
bohr, while it is $\lambda=2.2$ bohr at the  higher density of 
Au ($\overline{\rho}=0.098$
 bohr$^{-3}$). We may then roughly picture the response charge around Mg 
as a blob of Mg-density (low) electron gas surronded at close distance 
by blobs of Au-density (high) electron gas. The   screening 
wavelength around Mg  gets shorter (more akin to a high-density gas),
 approaching the efficiently-screening Au
sites:  we name this a variable-wavelength  Friedel
oscillation of  the core-hole screening density.

\vspace{-0.6cm}
\section{SUMMARY}

\vspace{-0.3cm}
Results of realistic, fully {\it ab initio} density-functional theory
calculations of core level shifts have been presented for the Mg 1$s$
and Au 4$f$ bulk-to-alloy shifts upon formation of the MgAu
intermetallic alloy. 
A large screening contribution was found for 
the Mg 1$s$ shift, whereas the same contribution is negligible for Au
4$f$. We observed unusual  features in the screening around
an excited Mg atom in MgAu, and  suggested
 a physical picture in terms of variable-wavelength Friedel 
oscillation around the Mg core hole, caused by the neighboring Au
atom in the alloy.

The {\it ab initio} treatment  provides
useful information about alloy core-level shifts,
which could not be obtained from experiment or models:
calculated core-level shift 
accurately reproduce experiment;
 theory  can quantify
 initial-state and screening contributions separately;
 initial-state shifts are found to be 
 compatible with a rigid-band model;
 screening densities at excited Au and Mg atoms enable to
 understand final-state contributions;
charge transfer is the central control parameter for the shifts.

%
%
%

%
 \end{document}